\documentclass[letterpaper, 10 pt, conference]{ieeeconf}  

\IEEEoverridecommandlockouts                              

\usepackage[pdftex]{graphicx}
\usepackage{import}
\usepackage{xcolor}
\usepackage{amsmath}

\title{\LARGE \bf
Introducing Risk Shadowing For Decisive and Comfortable \\ Behavior Planning
}

\author{Tim Puphal and Julian Eggert \\
Honda Research Institute (HRI) Europe, Carl-Legien-Str. 30, 63073 Offenbach, Germany \\ Email: {\tt\small \{firstname.lastname\}@honda-ri.de} 
}

\begin{document}

\maketitle
\thispagestyle{empty}
\pagestyle{empty}

\begin{abstract}
We consider the problem of group interactions in urban driving. State-of-the-art behavior planners for self-driving cars mostly consider each single agent-to-agent interaction separately in a cost function in order to find an optimal behavior for the ego agent, such as not colliding with any of the other agents. In this paper, we develop risk shadowing, a situation understanding method that allows us to go beyond single interactions by analyzing group interactions between three agents. Concretely, the presented method can find out which first other agent does not need to be considered in the behavior planner of an ego agent, because this first other agent cannot reach the ego agent due to a second other agent obstructing its way. In experiments, we show that using risk shadowing as an upstream filter module for a behavior planner allows to plan more decisive and comfortable driving strategies than state of the art, given that safety is ensured in these cases. The usability of the approach is demonstrated for different intersection scenarios and longitudinal driving.     
\end{abstract}

\section{Introduction}

Behavior planners for self-driving cars and driver support systems increasingly can tackle more complex driving situations. Here, most state-of-the-art behavior planners consider usually each agent separately to find a safe plan. Examples of behavior planners are search-based planners (such as A* search) \cite{sedighi2019}, trajectory optimization approaches \cite{ferguson2008} or cooperative planning methods \cite{puphal2019}. While considering each agent-to-agent interaction separately ensures safety, for decisive and comfortable behavior planning, interactions of groups should be considered for more proactive planning. 

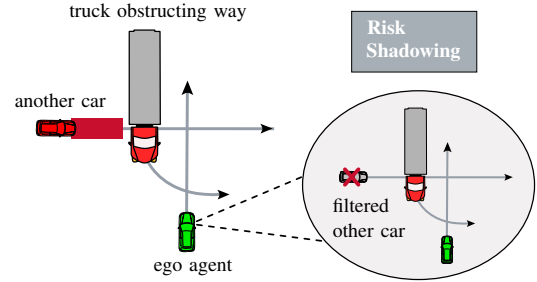
\begin{figure}[t!]
  \centering
  \vspace*{0.25cm}
  \resizebox{0.8\linewidth}{!}{
\begingroup%
  \makeatletter%
  \providecommand\color[2][]{%
    \errmessage{(Inkscape) Color is used for the text in Inkscape, but the package 'color.sty' is not loaded}%
    \renewcommand\color[2][]{}%
  }%
  \providecommand\transparent[1]{%
    \errmessage{(Inkscape) Transparency is used (non-zero) for the text in Inkscape, but the package 'transparent.sty' is not loaded}%
    \renewcommand\transparent[1]{}%
  }%
  \providecommand\rotatebox[2]{#2}%
  \newcommand*\fsize{\dimexpr\f@size pt\relax}%
  \newcommand*\lineheight[1]{\fontsize{\fsize}{#1\fsize}\selectfont}%
  \ifx\svgwidth\undefined%
    \setlength{\unitlength}{260.27194959bp}%
    \ifx\svgscale\undefined%
      \relax%
    \else%
      \setlength{\unitlength}{\unitlength * \real{\svgscale}}%
    \fi%
  \else%
    \setlength{\unitlength}{\svgwidth}%
  \fi%
  \global\let\svgwidth\undefined%
  \global\let\svgscale\undefined%
  \makeatother%
  \begin{picture}(1,0.52606168)%
    \lineheight{1}%
    \setlength\tabcolsep{0pt}%
    \put(0,0){\includegraphics[width=\unitlength,page=1]{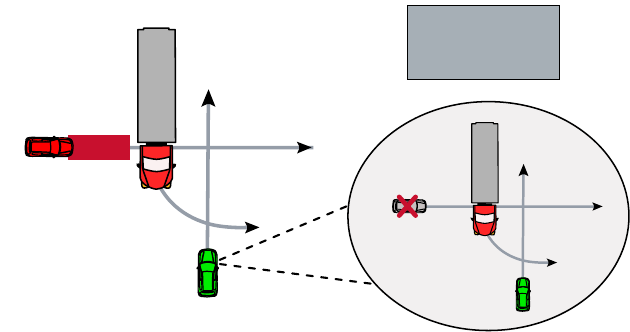}}%
    \put(0.10389429,0.5067157){\color[rgb]{0,0,0}\makebox(0,0)[lt]{\lineheight{1.25}\smash{\begin{tabular}[t]{l}truck obstructing way\end{tabular}}}}%
    \put(0.26648244,0.01734226){\color[rgb]{0,0,0}\makebox(0,0)[lt]{\lineheight{1.25}\smash{\begin{tabular}[t]{l}ego agent\end{tabular}}}}%
    \put(0.61077636,0.13098272){\color[rgb]{0,0,0}\makebox(0,0)[lt]{\lineheight{1.25}\smash{\begin{tabular}[t]{l}filtered \\other car\end{tabular}}}}%
    \put(0.67581252,0.42887936){\color[rgb]{1,1,1}\makebox(0,0)[lt]{\lineheight{1.25}\smash{\begin{tabular}[t]{l}\textbf{Shadowing}\end{tabular}}}}%
    \put(0.67673204,0.47327491){\color[rgb]{1,1,1}\makebox(0,0)[lt]{\lineheight{1.25}\smash{\begin{tabular}[t]{l}\textbf{Risk}\end{tabular}}}}%
    \put(-0.00117094,0.33394569){\color[rgb]{0,0,0}\makebox(0,0)[lt]{\lineheight{1.25}\smash{\begin{tabular}[t]{l}another car\end{tabular}}}}%
  \end{picture}%
\endgroup%
} 
  \vspace*{-0.2cm}
  \caption[]{The image shows an example driving situation in which another car cannot reach the ego agent due to the truck obstructing its way. In this paper, we propose risk shadowing which allows to filter other cars based on such group interactions and in turn allows the behavior planner of the ego agent to plan more decisive and comfortable behaviors.} 
  \label{fig:truck_example}
\end{figure}

Fig. \ref{fig:truck_example} illustrates a driving situation in which group interactions are important. The figure shows three agents at an intersection: an ego agent in green that intends to cross, while another car in red and another truck in red are approaching. Please note that the other car is driving at high velocity and the truck already drives on the intersection space. State-of-the-art behavior planners would consider both other agents separately in the motion planning problem and possibly recommend that the ego agent brakes for the fast other car. However, since the other car cannot reach the ego agent due to the truck obstructing its way, the car can be neglected from the perspective of the ego agent. In this paper, we  therefore present risk shadowing, a situation understanding method that allows us to analyze such group interactions. 

The novel risk shadowing approach consists of two different methods: a) a time-based risk model for predicting possible collisions and b) a reachability analysis based on geometrical constraints. Important is that both methods need to be applied from the perspective of each agent in the driving situation. In the example case of the figure, it has to be applied for the ego agent, the other car and the other truck. Since the computational effort required for group interactions increases quadratically with the number of involved agents, we propose to use a simple risk model, the closest encounter model, for this task.

\begin{figure*}[t!]
  \centering
  \vspace*{0.2cm}
  \resizebox{0.85\linewidth}{!}{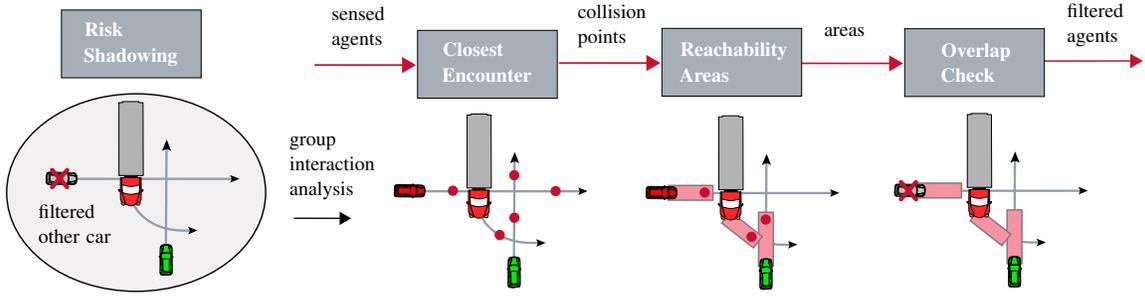}
  \vspace*{-0.15cm}
  \caption[]{The proposed risk shadowing initially utilizes the closest encounter model to predict future collision points, then determines reachability areas for the agents and lastly, performs an overlap check between the reachability areas. This allows the approach to analyze group interactions and to solve situations like the one from the introduction, filtering out the other car due to the truck obstructing its way.}
  \label{fig:risk_shadowing_concept}
\end{figure*}

The risk shadowing approach is finally used in the paper as an upstream filter module for a behavior planner. Agents which are filtered out by risk shadowing may be neglected in the planning step. In the example of Fig. \ref{fig:truck_example}, the planning system would thus not consider the other fast car due to the truck obstructing its way and recommend for the ego agent, for example, to keep its velocity and drive first through the intersection. The proposed system can better understand the driving situation and plans more decisive and comfortable behaviors than a system without risk shadowing. 

\subsection{Related Work}
\label{sec:rel}
To the knowledge of the authors, this is the first paper introducing a simple model-based group interaction analysis method for filtering in behavior planners. However, there are many works that are related to the proposed method.

First, while state-of-the-art behavior planners consider agents separately, recent learning-based prediction methods and neural planners may implicitly consider group interactions. For example, the authors of \cite{li2020} trained a model that predicts behavior trajectories with a graph neural network. If the driving data includes situations with group interactions, the model may implicitly learn these phenomena. Similarly, a reinforcement learning planner was trained in simulation in \cite{capasso2021} and might learn that some agents do not need to be considered because of the interaction with a third agent, and a lane change prediction based on three agents was learned in \cite{schmuedderich2015}. However, the generalizability to other driving situations is not ensured with these learning-based approaches.

Klingelschmitt et al. \cite{klingelschmitt2016} proposed a probabilistic situation recognition approach for multiple interacting agents. He shows that the consideration of multiple agents outperforms state-of-the-art methods.

Second, there is recently an extensive group of works that investigate reachability analysis. As an example, Althoff et al. published the work of \cite{pek2018} that allows to plan fail-safe behaviors based on reachable areas and the work of \cite{althoff2014} that allows to apply a safety check for given behavior planners. The proposed risk shadowing of this paper applies a similar reachability analysis. In contrast to these works, however, risk shadowing is used for filtering due to group interactions and not for safety verification of planners.

Third, there have been risk models applied previously for filtering in other works. The paper \cite{puphal2022} surveys different risk models for filtering agents in large intersection scenarios to reduce the computational cost for behavior planners. Besides the mentioned risk models in the paper, usable risk models for filtering are, for example, Responsibility Sensitive Safety (RSS) from \cite{shwartz2017} or acceleration-based methods, such as from \cite{wang2022}. All these risk models focus, however, only on single agent-to-agent interactions and not group interactions. 

\subsection{Contribution}
In summary, in this paper we introduce risk shadowing for decisive and comfortable behavior planning. The method applies a simple time-based risk model with a reachability analysis. Using this risk shadowing method as a filter module allows computationally cheaper planners and more intelligent driving. We will experimentally show the effectiveness of the resulting system. Here, the driving simulations are evaluated for different intersection scenarios and dynamic longitudinal driving, in which cars and trucks cross, follow or pass each other. 

The paper is structured as follows: Section II describes the implementation of the risk shadowing concept. We formulate the closest encounter model, compose reachability areas and calculate an overlap check to find out which agents can be neglected due to group interactions. Afterwards, Section III describes the system architecture of using risk shadowing as a filter and how we find an optimal plan with a behavior planner. Section IV demonstrates the results of the simulation experiments for multiple driving situations and Section V gives a summary and outlook for the paper. 

\section{Risk Shadowing Approach}
\label{sec:shadow_concept}

The overall framework of the risk shadowing approach is illustrated in Fig. \ref{fig:risk_shadowing_concept}. The figure explains the steps involved in risk shadowing and provides output examples for each of the steps. These examples are based on the driving situation from the introduction, which includes an ego car encountering another car that is obstructed by a truck. 

The proposed risk shadowing enables us to filter out the other car in the given driving situation using the following method. Initially, risk shadowing utilizes the closest encounter model to predict future potential collision points for the interactions between the agents. The collision points are visualized in the figure with red points. Then, our approach determines reachability areas for each agent, which are highlighted with red areas in the figure, based on these collision points. Lastly, risk shadowing performs an overlap check between the reachability areas. In the given driving situation, since the reachability area of the other car does not overlap with the ego car's area, risk shadowing can filter out the other car due to the truck obstructing its way, see the highlighted red cross above the other car.

Risk shadowing involves accordingly three steps. The first step is the risk model closest encounter, while the second and third step involve a reachability analysis, specifically determining reachability areas and applying an overlap check. In the next sections, we will describe each step of this framework in more detail.

\subsection{Closest Encounter Model}
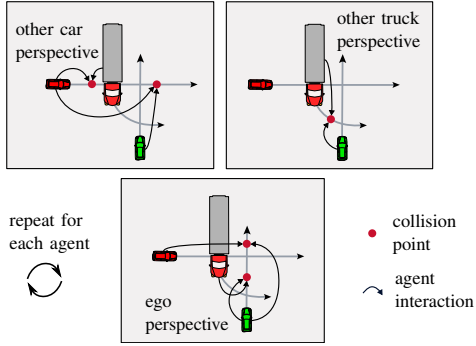
\begin{figure}[t!]
  \centering
  \vspace*{0.2cm}
  \resizebox{0.67\linewidth}{!}{
\begingroup%
  \makeatletter%
  \providecommand\color[2][]{%
    \errmessage{(Inkscape) Color is used for the text in Inkscape, but the package 'color.sty' is not loaded}%
    \renewcommand\color[2][]{}%
  }%
  \providecommand\transparent[1]{%
    \errmessage{(Inkscape) Transparency is used (non-zero) for the text in Inkscape, but the package 'transparent.sty' is not loaded}%
    \renewcommand\transparent[1]{}%
  }%
  \providecommand\rotatebox[2]{#2}%
  \newcommand*\fsize{\dimexpr\f@size pt\relax}%
  \newcommand*\lineheight[1]{\fontsize{\fsize}{#1\fsize}\selectfont}%
  \ifx\svgwidth\undefined%
    \setlength{\unitlength}{235.03952291bp}%
    \ifx\svgscale\undefined%
      \relax%
    \else%
      \setlength{\unitlength}{\unitlength * \real{\svgscale}}%
    \fi%
  \else%
    \setlength{\unitlength}{\svgwidth}%
  \fi%
  \global\let\svgwidth\undefined%
  \global\let\svgscale\undefined%
  \makeatother%
  \begin{picture}(1,0.78639358)%
    \lineheight{1}%
    \setlength\tabcolsep{0pt}%
    \put(0,0){\includegraphics[width=\unitlength,page=1]{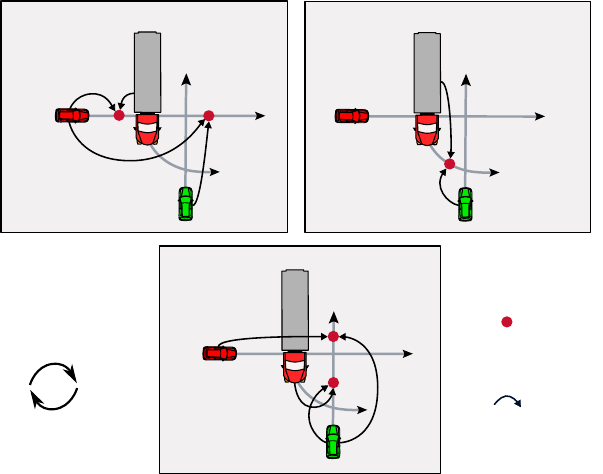}}%
    \put(0.88912961,0.26545102){\color[rgb]{0,0,0}\makebox(0,0)[lt]{\lineheight{1.25}\smash{\begin{tabular}[t]{l}collision \\point\end{tabular}}}}%
    \put(0.894264,0.13447999){\color[rgb]{0,0,0}\makebox(0,0)[lt]{\lineheight{1.25}\smash{\begin{tabular}[t]{l}agent \\interaction\end{tabular}}}}%
    \put(0.02274722,0.70859779){\color[rgb]{0,0,0}\makebox(0,0)[lt]{\lineheight{1.25}\smash{\begin{tabular}[t]{l}other car \\perspective\end{tabular}}}}%
    \put(0.76107532,0.73386674){\color[rgb]{0,0,0}\makebox(0,0)[lt]{\lineheight{1.25}\smash{\begin{tabular}[t]{l}other truck \\perspective\end{tabular}}}}%
    \put(0.32092611,0.08154847){\color[rgb]{0,0,0}\makebox(0,0)[lt]{\lineheight{1.25}\smash{\begin{tabular}[t]{l}ego \\perspective\end{tabular}}}}%
    \put(0.0086249,0.2692757){\color[rgb]{0,0,0}\makebox(0,0)[lt]{\lineheight{1.25}\smash{\begin{tabular}[t]{l}repeat for\\each agent\end{tabular}}}}%
  \end{picture}%
\endgroup%
}
  \vspace*{-0.15cm}
  \caption[]{In the closest encounter model, the collision points are predicted from the perspective of each agent to analyze group interactions. The model predicts the agents' motion and computes the points of collision.}  
  \label{fig:closest_encounter}
\end{figure}

The closest encounter model is used to find out future collision points between the agents in the driving situation. Fig. \ref{fig:closest_encounter} shows this process of predicting collision points in detail. An important aspect is that the model is required to be applied from each agent's perspective. Therefore, the figure shows for the driving situation of three agents, the collision points from the perspective of the other car, the other truck, and the ego car. Applying the model for each agent allows to consider group interactions between the agents. 

As an example, the determination of a collision point between two agents is described in the following. The closest encounter model predicts the agents' motion along their given driving paths with constant velocity. This results in, for example, two position sequences $\textbf{x}_1(s)$ and $\textbf{x}_2(s)$ with the considered agent $1$, another agent $2$ and the future time~$s$. The distance between the agents is subsequently computed for each timestep with the formula
\begin{align}
d(s)= ||\textbf{x}_2(s) - \textbf{x}_1(s)||.\text{\footnotemark}
\end{align}
\footnotetext{Please note that in the current implementation, the agents' vehicle shapes are additionally considered for more accurate distance values.}
\vspace*{-0.45cm}

\noindent The closest encounter model is then composed of the three variables: distance of closest encounter (DCE), time of closest encounter (TCE) and point of closest encounter (PCE). These variables can be derived based on the distance over the future time $d(s)$ according to
\begin{align}
\text{DCE} &= \text{min}_{s} \{d(s)\}, \\
\text{TCE} &= \text{argmin}_{s} \{d(s)\}, \\
\text{PCE} &= \textbf{x}_{1}(s=\text{TCE}).
\end{align}
Here, a collision point requires that DCE needs to be smaller than a threshold $d_{\text{thr}}$ and we finally write for the collision point $\textbf{x}_{\text{coll}}$ the condition
\begin{align}
\text{if DCE} < d_{\text{thr}} \hspace{-0.085cm}: \hspace{0.05cm} \textbf{x}_{\text{coll}} = \text{PCE}.
\end{align}

Fig. \ref{fig:closest_encounter} illustrates on the top left box, amongst others, these collision points from the other car's perspective. These are the first collision point with the other truck and the second collision point with the ego car. We specify multiple collision points with $\textbf{x}_{\text{coll},j}$, whereby $j$ represents the index of the given collision point. The process is repeated for the truck and the ego car, as can be seen in the figure.

\subsection{Reachability Areas}

\begin{figure}[t!]
  \centering
  \vspace*{0.2cm}
  \resizebox{0.67\linewidth}{!}{
\begingroup%
  \makeatletter%
  \providecommand\color[2][]{%
    \errmessage{(Inkscape) Color is used for the text in Inkscape, but the package 'color.sty' is not loaded}%
    \renewcommand\color[2][]{}%
  }%
  \providecommand\transparent[1]{%
    \errmessage{(Inkscape) Transparency is used (non-zero) for the text in Inkscape, but the package 'transparent.sty' is not loaded}%
    \renewcommand\transparent[1]{}%
  }%
  \providecommand\rotatebox[2]{#2}%
  \newcommand*\fsize{\dimexpr\f@size pt\relax}%
  \newcommand*\lineheight[1]{\fontsize{\fsize}{#1\fsize}\selectfont}%
  \ifx\svgwidth\undefined%
    \setlength{\unitlength}{235.03952291bp}%
    \ifx\svgscale\undefined%
      \relax%
    \else%
      \setlength{\unitlength}{\unitlength * \real{\svgscale}}%
    \fi%
  \else%
    \setlength{\unitlength}{\svgwidth}%
  \fi%
  \global\let\svgwidth\undefined%
  \global\let\svgscale\undefined%
  \makeatother%
  \begin{picture}(1,0.78639358)%
    \lineheight{1}%
    \setlength\tabcolsep{0pt}%
    \put(0,0){\includegraphics[width=\unitlength,page=1]{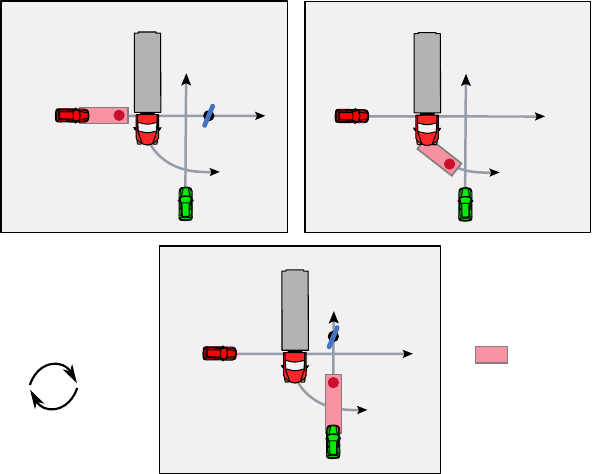}}%
    \put(0.02274704,0.70859377){\color[rgb]{0,0,0}\makebox(0,0)[lt]{\lineheight{1.25}\smash{\begin{tabular}[t]{l}other car \\perspective\end{tabular}}}}%
    \put(0.76107485,0.73386271){\color[rgb]{0,0,0}\makebox(0,0)[lt]{\lineheight{1.25}\smash{\begin{tabular}[t]{l}other truck \\perspective\end{tabular}}}}%
    \put(0.32092593,0.08154449){\color[rgb]{0,0,0}\makebox(0,0)[lt]{\lineheight{1.25}\smash{\begin{tabular}[t]{l}ego \\perspective\end{tabular}}}}%
    \put(0.00862461,0.26927168){\color[rgb]{0,0,0}\makebox(0,0)[lt]{\lineheight{1.25}\smash{\begin{tabular}[t]{l}repeat for\\each agent\end{tabular}}}}%
    \put(0.87825197,0.2047396){\color[rgb]{0,0,0}\makebox(0,0)[lt]{\lineheight{1.25}\smash{\begin{tabular}[t]{l}reachability\\area\end{tabular}}}}%
  \end{picture}%
\endgroup%
}
  \vspace*{-0.15cm}
  \caption[]{In the reachability area determination, the reachability areas are calculated by finding the closest collision point along the considered agent's driving path. Other collision points are disregarded.} 
  \label{fig:reachable_area}
\end{figure}
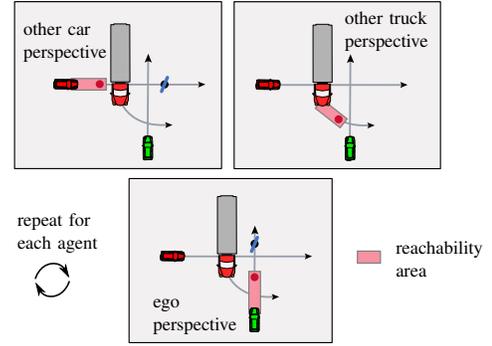

Once the collision points are determined for every interaction in the group of the driving situation, the reachability areas can be determined in the next step. 
Fig. \ref{fig:reachable_area} illustrates the process of the reachability area determination. We define the reachability area (RA) of an agent to start at the agent's current position and to end at the closest collision point along the agent's driving path. An agent will not be able to drive further than this first collision point. To simplify the problem, we thus project the agent's position $\textbf{x}_{\text{agent}}$ and collision point results $\textbf{x}_{\text{coll},j}$ onto the respective path to obtain longitudinal positions along the path $l_{\text{agent}}$ and $l_{\text{coll},j}$. A reachability area can consequently be calculated by 
\begin{align}
\text{RA} &= [l_{\text{agent}}, \hspace{0.05cm} \text{min}_j\{l_{\text{coll},j}\}].
\end{align}
In the equation, the reachability area is modeled as a one-dimensional line. The agents' width is added afterwards for a two-dimensional representation. 

Fig. \ref{fig:reachable_area} shows the described process of the reachability area determination from the perspective of each agent. A reachability area is determined for the other car, other truck and ego agent. For example, the other car's reachability area starts at the car's current position and ends at the first collision point with the truck. The second collision point with the ego car comes after the first collision point and therefore remains disregarded, which is illustrated with a blue line that crosses out the point. Risk shadowing can proceed now to the final step, which is the overlap check.

\subsection{Overlap Check}
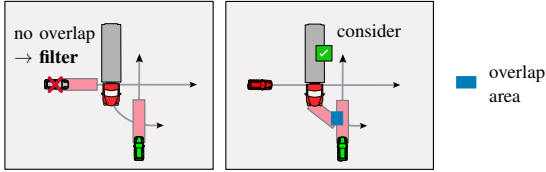
\begin{figure}[t!]
  \centering
  \vspace*{0.2cm}
  \resizebox{0.73\linewidth}{!}{
\begingroup%
  \makeatletter%
  \providecommand\color[2][]{%
    \errmessage{(Inkscape) Color is used for the text in Inkscape, but the package 'color.sty' is not loaded}%
    \renewcommand\color[2][]{}%
  }%
  \providecommand\transparent[1]{%
    \errmessage{(Inkscape) Transparency is used (non-zero) for the text in Inkscape, but the package 'transparent.sty' is not loaded}%
    \renewcommand\transparent[1]{}%
  }%
  \providecommand\rotatebox[2]{#2}%
  \newcommand*\fsize{\dimexpr\f@size pt\relax}%
  \newcommand*\lineheight[1]{\fontsize{\fsize}{#1\fsize}\selectfont}%
  \ifx\svgwidth\undefined%
    \setlength{\unitlength}{250.9954709bp}%
    \ifx\svgscale\undefined%
      \relax%
    \else%
      \setlength{\unitlength}{\unitlength * \real{\svgscale}}%
    \fi%
  \else%
    \setlength{\unitlength}{\svgwidth}%
  \fi%
  \global\let\svgwidth\undefined%
  \global\let\svgscale\undefined%
  \makeatother%
  \begin{picture}(1,0.35681958)%
    \lineheight{1}%
    \setlength\tabcolsep{0pt}%
    \put(0,0){\includegraphics[width=\unitlength,page=1]{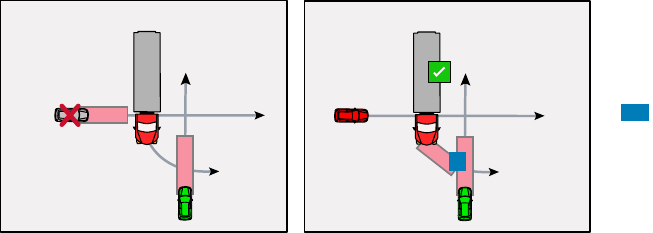}}%
    \put(0.70205,0.28033295){\color[rgb]{0,0,0}\makebox(0,0)[lt]{\lineheight{1.25}\smash{\begin{tabular}[t]{l}consider\end{tabular}}}}%
    \put(0.02046134,0.27570699){\color[rgb]{0,0,0}\makebox(0,0)[lt]{\lineheight{1.25}\smash{\begin{tabular}[t]{l}no overlap\\$\rightarrow$ \textbf{filter}\end{tabular}}}}%
    \put(1.0226761,0.1934807){\color[rgb]{0,0,0}\makebox(0,0)[lt]{\lineheight{1.25}\smash{\begin{tabular}[t]{l}overlap\\area\end{tabular}}}}%
  \end{picture}%
\endgroup%
}
  \vspace*{-0.12cm}
  \caption[]{For the ego agent, an overlap check is applied. If the reachability area of the ego agent and the reachability area of another agent do not overlap, the other agent is filtered out.}  \label{fig:overlap_check}
\end{figure}

In the final step of risk shadowing, the reachability areas of the ego agent and other agents are checked for overlaps. Fig. \ref{fig:overlap_check} shows this check for the ego car in the driving situation of the introduction example. 

Based on the reachability areas, we can conclude whether another agent can be filtered out or not. In particular, if the reachability area of the ego agent and of another agent do not overlap, the agent may be filtered out. We therefore define for another agent $i$ that if the condition 
\begin{align}
\text{RA}_{\text{ego}} \cap \text{RA}_{\text{other}, i} = \emptyset   
\end{align}
holds, then the other agent $i$ is filtered out. The agent can 

\noindent be safely neglected because it cannot reach the ego agent in the current driving constellation.

As Fig. \ref{fig:overlap_check} depicts on the left box, the reachability area of the ego car does not overlap with the reachability area of the other car because the other car's area already ends before the truck. The other car can thus safely be filtered out and does not need to be considered for an upcoming behavior planner. The risk shadowing approach found out that the other car is obstructed by the truck and considers this group influence for the ego car. On the other hand, the ego car's reachability area overlaps with the other truck's reachability area, see the right box. This is highlighted with a blue overlap area. The truck is thus considered according to risk shadowing.

\section{Planning with Filter}
\label{sec:planning_with_filter}

In the last section, the proposed risk shadowing approach of this paper was explained based on the introduction example of three agents. Since group interaction analysis requires some steps to be repeated from the perspective of each agent, the time complexity is $O(n^2)$, with $n$ being the number of agents. For this reason, in this paper, the models were chosen to be simple. The closest encounter was chosen as a simple risk model for the collision predictions and the reachability area determination was done with one-dimensional lines. Risk shadowing can efficiently filter out agents that cannot reach the ego agent based on group influences. 

Fig. \ref{fig:risk_shadowing_in_planner} shows now the usage of the risk shadowing approach as an upstream filter module for a behavior planning module. As an input, risk shadowing requires the information from sensors, e.g., from cameras or lidars, for all sensed agents in the driving situation. The information includes the position, velocity, and their future driving paths from, e.g., map data. As an output, risk shadowing gives a reduced set of agents that does not include the filtered out agents. Only relevant other agents are considered.

The behavior planner utilizes the reduced set of agents as an input and plans an optimal behavior for the ego agent. This simplifies the behavior planning problem. In this paper, we employ Risk Maps \cite{damerow2014} for the behavior planning module as explained in detail in \cite{puphal_journal2022}. Risk Maps generates here multiple ego behaviors and checks their risk, utility and comfort for choosing an optimal behavior. 
The found behavior is finally executed. In the following, we will shortly explain Risk Maps to conclude the methods sections.

\subsection{Risk Maps Planner}
\begin{figure}[t!]
  \centering
  \vspace*{0.25cm}
  \resizebox{0.9\linewidth}{!}{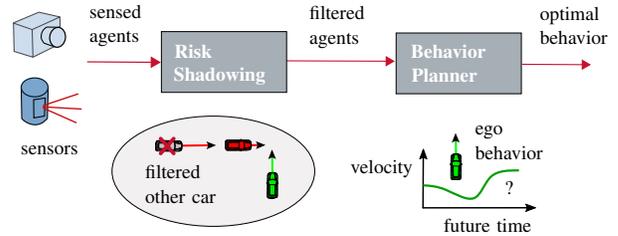}
  \vspace*{-0.07cm}
  \caption[]{\hspace{0.07cm} Risk shadowing is used in this paper as an upstream filter module for the behavior planner Risk Maps. From the sensed agents, agents are filtered out, and Risk Maps only uses the remaining set of agents to find an optimal behavior.} 
  \label{fig:risk_shadowing_in_planner}
\end{figure}

Risk Maps \cite{damerow2014, puphal_journal2022} models the driving behavior of the ego agent with taken velocity profiles over the future time, which follow the given driving path. Accordingly, we solve behavior planning with a velocity planner, as can be seen with the velocity plot at the bottom right of Fig. \ref{fig:risk_shadowing_in_planner}. 

Risk Maps generates first for the ego agent several potential velocity profiles $v^h$ that are composed of $h$ different acceleration and deceleration behaviors. The optimal behavior is then the velocity profile $v_{\text{opt}}$ among these generated profiles that has the lowest driving costs $C$. The costs $C$ consists here of future risks $R$, utility $U$ and comfort costs $O$. We therefore write
\begin{align}
v_{\text{opt}} = \text{argmin}_{h} \{C(v^h)\}, \\
\text{with } C(v^h) = R(v^h) - U(v^h) &+ O(v^h), 
\end{align} 
for the planning formulation of Risk Maps. As an example, an optimal behavior for the ego agent could be to brake in order to allow another agent to pass and then to keep constant velocity for the rest of the planning time.

The used risks in the cost formulation of Risk Maps are probabilistic and include, on the one hand, risks for the ego agent to collide with other agents, and, on the other hand, risks for the ego agent to lose control in sharp turns. While Risk Maps targets to minimize these risks, at the same time, the behavior planner also maximizes utility and minimizes comfort costs. Utility is therein represented by the traveled distance towards the ego agent's goal point and comfort costs penalize sudden changes in the velocity profile, such as high acceleration or jerk values. 

For more details about the velocity profile modeling, the cost formulation and the used parameters in Risk Maps, we refer the reader to the work of \cite{puphal_journal2022}. At this point, however, we want to note that other behavior planners can be used as well. The main contribution of this paper is the risk shadowing approach that helps behavior planners to reduce the driving situation's complexity.

\begin{figure*}[t!]
  \centering
  \vspace*{0.15cm}
  \resizebox{0.92\linewidth}{!}{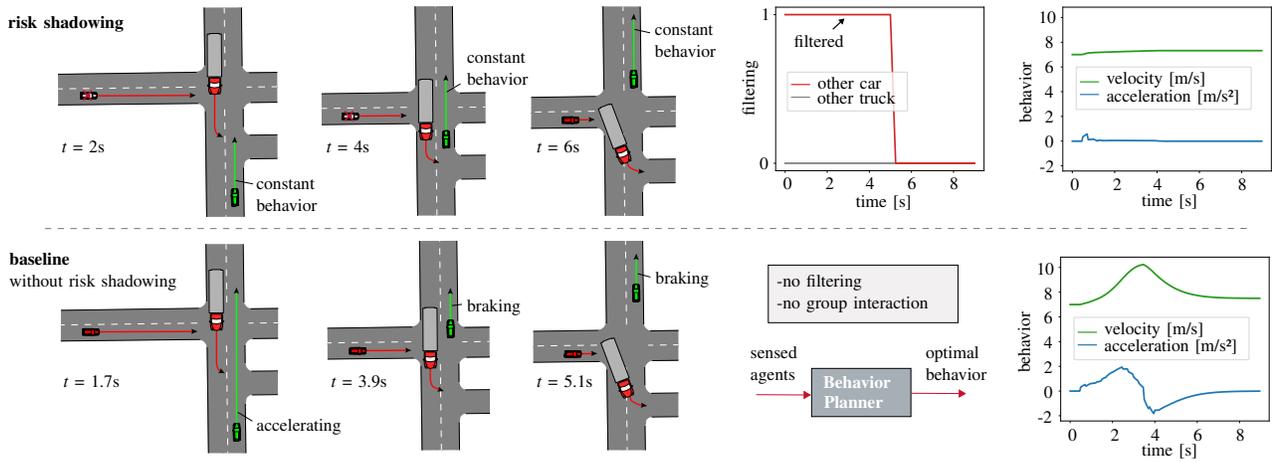}
  \vspace{-0.13cm}
  \caption[]{The image shows risk shadowing and Risk Maps being applied on the introduction example that includes another car obstructed by a truck. With risk shadowing, the other car is filtered out and the planner finds a decisive and comfortable behavior of crossing the intersection with constant velocity. In contrast, the baseline system without risk shadowing accelerates and brakes in response to the other car.} 
  \label{fig:filtering_intro_scenario}
\end{figure*}

\section{Experiments}
\label{sec:exp}
In this section, we will finally present the experiments of this paper that evaluate the performance and robustness of the risk shadowing approach. In the experiments, we applied risk shadowing as an upstream filter module for a behavior planner as described earlier. For this purpose, simulations were carried out using a traffic simulator, where the ego agent followed the planned behavior generated by the combined system of risk shadowing and Risk Maps, and other agents drove with fixed behaviors. To model the behavior of the agents, we used here kinematic models that updated their positions along the respective driving paths.

The section is divided into three parts. First, we will show simulations of the introduction example including another car that is obstructed by a truck. We analyze the output of risk shadowing and of the behavior planner in order to show the performance of the approach. Second, different variations for the intersection scenario are shown that involve three agents interacting with each other and we highlight different filter and non-filter examples to analyze the robustness of risk shadowing. Third, we will discuss the limitations of the current implementation of risk shadowing. 

In order to show the advantages of the risk shadowing approach, we will compare the novel system of risk shadowing and Risk Maps with a simple system using only the behavior planner Risk Maps. We therefore label in the experiments the novel system as "risk shadowing" and the simple system as "baseline".

\subsection{Introduction Example}

The improvement of using risk shadowing with a behavior planner for the driving situation of the introduction example is depicted in Fig. \ref{fig:filtering_intro_scenario}. The top portion of the figure shows the results obtained by applying "risk shadowing" and the bottom portion the results obtained by applying the "baseline" without risk shadowing. 

We now focus on the figure's top portion, which illustrates birds-eye-views of the driving situation for three timesteps (left). The ego agent in green is crossing an intersection, while another car in red and another truck in red are present. By applying risk shadowing, the complexity of the driving situation is simplified. Due to the truck obstructing the way for the other car, the car is filtered out. The behavior planner Risk Maps thus finds that the optimal behavior is to maintain a constant velocity and the ego agent crosses the intersection. The filtering and behavior graphs (right) demonstrate this circumstance. The other car is filtered out until around the simulation time of $5$ seconds in the filtering graph and the behavior graph contains constant velocity and no acceleration for the ego agent over the complete simulation.

In contrast, at the bottom, the figure presents the same simulation using the "baseline" system. The system consists only of the behavior planner Risk Maps. Here, there is no filtering and no group interaction analysis applied. 

The other car, which is approaching the intersection, has  a strong influence on the ego car and the planner recommends to initially accelerate in order to avoid the other car, and, after crossing the intersection, to brake for returning to the original velocity (see the birds-eye views of the driving situation). The planner is unaware that the truck is obstructing the way for the other car and minimizes this separate agent-to-agent collision risk. The behavior graph demonstrates that the ego agent accelerates from $0$ to $3$ seconds and then decelerates from $3$ to $8$ seconds in the simulation.

Overall, the introduction example has shown that applying risk shadowing for intersection scenarios with group interactions allows the behavior planner to find a more decisive and comfortable behavior. The planner recommends to keep the current velocity instead of unnecessarily changing its behavior. 

\begin{figure}[t!]
  \centering
  \vspace*{0.15cm}
  \hspace*{0.07cm}
  \resizebox{0.81\linewidth}{!}{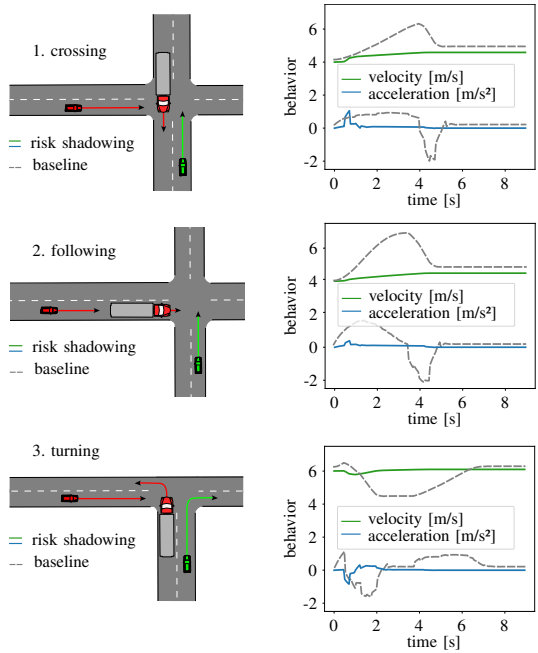}
  \caption[]{The improvement of risk shadowing for behavior planning is shown for crossing, following and turning scenarios. A more decisive and comfortable behavior is also achieved in these scenarios.} 
  \label{fig:variety_intersections}
\end{figure}

\subsection{Further Scenarios}
Fig. \ref{fig:variety_intersections} shows further examples with three agents interacting on an intersection exhibiting benefits of risk shadowing. In particular, risk shadowing is analyzed for crossing, following and turning intersection scenarios which cover a variety of possible interactions. 

The figure depicts for each of the scenarios the birds-eye-view of the driving situation for one timestep during the simulation on the left and the behavior graph of the ego agent for the "risk shadowing" approach and the "baseline" on the right. In the scenarios, risk shadowing allows to filter out the other car because of the obstruction from the truck. The planner finds decisive and comfortable behaviors by crossing the intersection using constant behavior. In comparison, the baseline system without risk shadowing overestimates the risk from the other car in the scenarios and recommends the ego agent to change its behavior by strongly accelerating or braking. The baseline system hence assesses the situation wrong for group interactions. 

\begin{figure}[t!]
  \centering
  \vspace*{0.2cm}
  \resizebox{0.92\linewidth}{!}{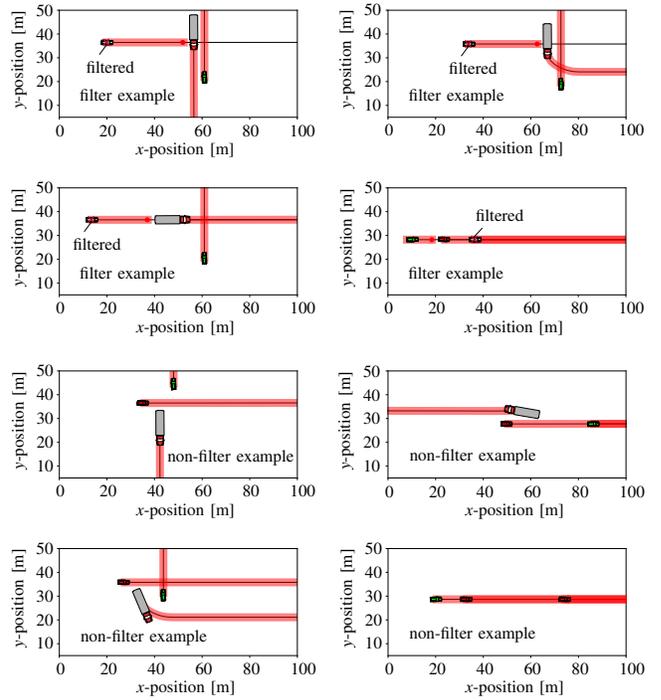}
  \caption[]{The determined reachability areas of the risk shadowing approach are visualized for different filter examples and non-filter examples. Besides intersection scenarios, the robustness of risk shadowing was tested for longitudinal following.}
  \label{fig:risk_shadowing_examples}
\end{figure}

Furthermore, to give insights into the calculation internals of the risk shadowing approach, the determined reachability areas of each agent for the different scenarios are visualized in Fig. \ref{fig:risk_shadowing_examples}. The figure presents on the top four filter examples where another car is filtered out because of a third vehicle obstructing its way. At the bottom of the figure, in contrast, four non-filter examples, in which reachability areas either have an overlap or the agents have already passed each other, are shown. In the scenarios, risk shadowing correctly assesses the group interactions. 

We want to highlight the presented longitudinal following scenarios in the figure. As can be seen in the filter examples (second row on the right), another car is filtered out by risk shadowing if the ego agent cannot reach this car because of a third car in between. This is true when the car in between, for example, brakes strongly. The non-filter example scenario of longitudinal following (fourth row on the right) shows that this car should, however, not be filtered out if the agents drive all with similar velocities.

\subsection{Discussion}
In total, the experiments have shown that the risk shadowing approach can be applied to different driving situations and allows to filter out other agents because of group interactions. However, limitations of the approach are discussed in the following. 

In the intersection experiments, we analyzed group interactions where one agent is driving close to or on the intersection space. The current implementation of risk shadowing does not filter out another agent based on group interactions, if all the agents in the group are distant from the intersection space. Analyzing such group interactions can also result in a more intelligent behavior for the ego agent and should be investigated. 

In addition, in both, the shown intersection scenarios and longitudinal following examples, the agents were assumed to have one driving path. It is possible to consider multiple path options for one agent by applying risk shadowing on each path and only filtering out another agent if it is obstructed on all its paths. Nevertheless, this scales unfavorably if the path options are numerous for each agent. Further heuristics should be investigated and tested to filter out agents in these situations.

The last limitation of the current risk shadowing approach is the assumption of constant velocity in the used closest encounter model. This assumption may not hold true when agents are exerting strong accelerations and decelerations, which can result in inaccurate predictions. To improve risk shadowing, safety margins may be added in the reachability areas to account for errors in the prediction and further tests that parametrize these safety margins are needed.

\section{Conclusion and Outlook}
In summary, in this paper, we proposed a situation understanding method called risk shadowing that allows to analyze group interactions. The approach implements a) a time-based risk model and b) a reachability analysis to find out if another agent can be filtered out and neglected in a behavior planner of an ego agent, because the way of this first other agent is obstructed by a further second other agent. Risk shadowing is implemented here in a computationally efficient manner by using, amongst others, the simple closest encounter model for the risk model. 

In experiments, we finally showed that this risk shadowing allows to reduce the situation complexity for different driving situations. The combined system of risk shadowing with the behavior planner Risk Maps allowed to plan more decisive and comfortable driving strategies than state of the art not using risk shadowing. While the combined system could drive the ego agent with constant behavior through intersections, the baseline system recommended unnecessary accelerating and braking behaviors. The analyzed driving situations included intersection scenarios where agents were crossing, following and turning as well as examples of pure longitudinal following.

A remaining limitation of the risk shadowing method for intersection scenarios is that the method can only be applied if one of the agents in the group is already located close to the intersection space. In future work, we therefore would like to investigate a more continuous than discrete risk model for group interactions. This would allow us to filter out agents distant to the intersection space and cover further phenomena of group interactions. Such a filter could be interesting, for example, for future self-driving cars.

\addtolength{\textheight}{-12cm}   

\vspace{0.13cm}
\bibliographystyle{IEEEtran}
\bibliography{bib}
\end{document}